Review article

Luigi Consolino*, Francesco Cappelli, Mario Siciliani de Cumis and Paolo De Natale

# QCL-based frequency metrology from the mid-infrared to the THz range: a review

**Abstract:** Quantum cascade lasers (QCLs) are becoming a key tool for plenty of applications, from the mid-infrared (mid-IR) to the THz range. Progress in related areas, such as the development of ultra-low-loss crystalline microresonators, optical frequency standards, and optical fiber networks for time and frequency dissemination, is paving the way for unprecedented applications in many fields. For most demanding applications, a thorough control of QCLs emission must be achieved. In the last few years, QCLs' unique spectral features have been unveiled, while multifrequency QCLs have been demonstrated. Ultra-narrow frequency linewidths are necessary for metrological applications, ranging from cold molecules interaction and ultra-high sensitivity spectroscopy to infrared/THz metrology. A review of the present status of research in this field is presented, with a view of perspectives and future applications.

# 1 Introduction

One of the most challenging photonic endeavors for the last decades is the realization of metrological, tunable, and spectrally pure laser sources in the mid-infrared (mid-IR) and in the THz range. Work in these regions is crucial for high precision molecular spectroscopy; in the so-called fingerprint region, the line strength of absorption lines is generally larger with respect to the near-infrared (near-IR), while Doppler-linewidths linearly decrease with increasing wavelength.

Performing molecular precision spectroscopy enables very attractive fundamental physics research about, for example, time variation of fundamental constants ($\alpha$, $m_p/m_e$, …), electric dipole moment of the electron, and parity violation in molecules [1–5]. At the same time, mid-IR/THz spectroscopy has a considerable impact in real-world applications spanning from environmental monitoring to health, from imaging to security [6–15].

Efficient nonlinear generation of mid-IR and THz radiation with excellent spectral properties, performed since the 1980s of the XXth century, opened the way to metrological research in these spectral regions [16, 17]. However, the optical power available for experiments and the complexity of such setups were not suitable for compact and practical applications.

A crucial step towards the development of high-resolution spectroscopy and metrology in these regions was the invention of quantum cascade lasers (QCLs) [18], one of the most significant developments for semiconductor physics for the last 25 years. Since 1994, compact, powerful, stable, tunable, laser sources have become available in these spectral regions. This invention entailed the development of technology and components in the mid-IR and in the THz, which have speeded up research and applications. QCLs covering the wide mid-IR and THz regions of the electromagnetic spectrum (with a gap between ~6 and 10 THz) have allowed to fill the technological gap, as compared to other regions, due to the limitations of other kinds of sources, like OPOs, $CO_2$, difference frequency generation (DFG) setups, lead salt lasers, or molecular far-infrared lasers, thus significantly extending the range of use of compact semiconductor sources.

Since their first demonstration, QCLs operating in the mid-IR have undergone an impressive development, achieving high performance levels. For example, the maximum operating temperature can be even higher than room temperature (RT) in a wide range of wavelengths (5–12 μm) [19]. Significantly, RT continuous-wave (CW) single-mode operation has been demonstrated [20]. Multi-watt output power, continuous wave, RT devices operating

*Corresponding author: Luigi Consolino, Istituto Nazionale di Ottica, CNR-INO, Largo E. Fermi, 6 – 50125 Firenze, Italy, e-mail: luigi.consolino@ino.it
**Francesco Cappelli and Paolo De Natale:** Istituto Nazionale di Ottica, CNR-INO, Largo E. Fermi, 6 – 50125 Firenze, Italy
**Mario Siciliani de Cumis:** Agenzia Spaziale Italiana, ASI, Contrada Terlecchia snc, Matera, Italy

across the mid-IR, with wall plug efficiencies larger than 50%, have been recently reported [21, 22], with impressive performance in terms of spectral coverage (~3–25 μm) and tunability range [23]. QCLs with new promising material systems have been recently demonstrated to work up to 400 K at wavelengths within the first atmospheric window (3–5 μm) [24, 25]. In 2002, the spectral coverage of QCLs was extended to the far-IR [26], where efficient and miniaturized sources operating in the 1.2–4.9-THz window have now been successfully developed in either single plasmon or double-metal waveguide configuration [26]. Currently, the challenge is to increase the operating temperature of THz QCLs and also, from a metrological point of view, enhance the performance of such devices, narrowing their emission linewidth, referencing them to a stable frequency standard, and studying quantum properties in order to push forward fundamental research.

## 2 Quantum cascade lasers

QCL emission is based on intersubband transitions, first proposed in 1971 by Kazarinov and Suris [27]. The development of growth techniques, such as molecular beam epitaxy (MBE – see Figure 1 and ref. [29]), enabling unprecedented control on semiconductor layers thickness, opened the way to the design of new materials by semiconductor bandgap engineering [30]. In this context, the possibility of using heterostructures to modulate the bandgap, creating sharp discontinuities in the conduction and valence bands, has enabled the investigation of new phenomena and devices.

Actually, QCLs [18, 28, 31] can be considered as the primary achievement in electronic band structure engineering, showing how artificial materials can be created through quantum design to have tailor-made properties. QCLs are unipolar devices exploiting optical transitions between electronic states (conduction subbands) created by the spatial confinement of the electrons in semiconductor quantum wells. QCLs have a ground-breaking design based on the engineering of electronic wavefunctions on a nanometer scale (see Figure 2).

As far as the macroscopic properties of materials are defined by their electronic structure, the QCL is based on an artificial nano-material. The extreme precision of the material growth that is required to get the proper operation properties, combined with the large number of layers and the complexity of the structure, gives an impressive demonstration of the capabilities offered by bandgap engineering, with still much potential to explore novel

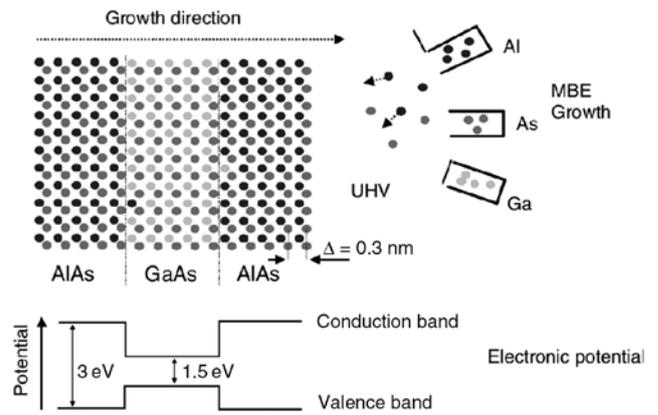

**Figure 1:** Representation of molecular beam epitaxy growth (MBE) of a GaAs/AlAs structure, together with the related electronic potential.
Adapted with permission from [28], copyright 2013.

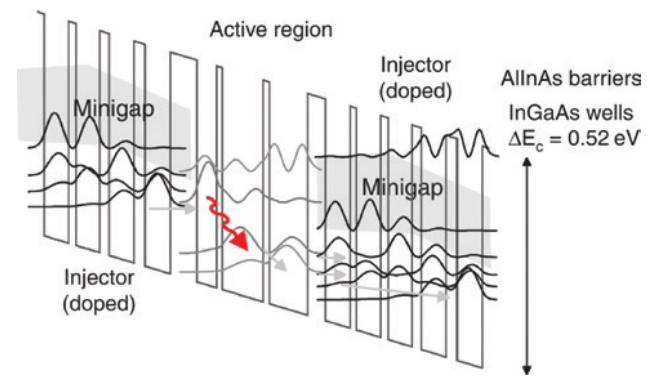

**Figure 2:** Schematic of the band structure of a QCL.
The scheme shows a three-quantum-wells active region design for λ = 10.3 μm operation, preceded and followed by injector stages. Together with the semiconductor superlattice, the square moduli of the electronic wavefunctions are depicted. The overall slope is given by the externally applied electric field. The gray straight arrows denote electrons flow, while the red wavy arrow denotes the lasing transition. Adapted with permission from [28], copyright 2013.

quantum physical parameters and endow QCLs with brand new capabilities.

A peculiar feature of QCLs is the possibility to engineer the emission frequency over a large bandwidth with the same semiconductor material by changing the size of quantum wells to vary the energy separation of electronic states. This distinctive feature, together with the unipolar nature of the carriers transport and the peculiar shape of the density of states, enables performances totally different from those of bipolar lasers, which have an emission wavelength constrained by the material bandgap and a gain that is strongly temperature-dependent. Moreover, in contrast to conventional interband semiconductor lasers, in QCLs, the gain linewidth depends only indirectly on the temperature, and the optical gain is not limited by the joint density of states [28, 32]. This leads to the absence of gain saturation when electron and hole quasi-Fermi levels are well within conduction and valence bands. The gain is therefore only limited by the amount of current that can be injected in the structure to sustain electron population inversion. In addition, the multistage cascaded geometry allows for electron recycling, so that each electron injected may generate a number of photons equal to the number of stages. The cascade geometry has the significant advantage that a uniform gain across the active region is limited by the ratio of the effective transit times between the wells, including capture of the slower carrier and recombination times. The number of stages is mostly limited by the ratio between the effective width of the optical mode and the length of an individual stage.

The tremendous progress these sources have undergone in the last decades has been possible, thanks to a thorough characterization of these devices, e.g. unveiling their unique spectral features. Indeed, extremely narrow frequency linewidths, at tens of Hz or below, are necessary for demanding applications ranging from cold molecules interaction [33, 34], ultra-high sensitivity spectroscopy [35], and trace-gas sensing [15, 36–39] to infrared/THz metrology [40, 41]. Towards these goals, complete characterization and control of the emission of QCLs are necessary. In fact, although QCLs have shown extremely high spectral purity both in the mid-IR and in the THz domain, a crucial step towards an extensive use of QCLs for demanding spectroscopic and metrological applications is the development of techniques enabling not only the narrowing of QCLs emission down to the kilohertz level but also its referencing to a stable frequency standard. These opportunities will be discussed in details not only for already available mid-IR devices and THz QCLs but also for new generation sources that are emerging as the new frontier of unipolar devices, including mid-IR QCL combs and RT THz QCLs.

## 3 Mid-IR QCLs

In the mid-IR region, intense and narrow (i.e. with a low frequency-noise) laser sources are needed to perform high-sensitivity and high-resolution sub-Doppler spectroscopy. Moreover, if also a high accuracy is required, i.e. control on systematic uncertainties, an absolute reference for frequencies is needed. QCLs are ideal candidates for this role, as their intrinsic linewidth is comparable to the natural linewidth of molecular transitions (tens to hundreds of Hz) [42, 43], and the emitted radiation intensity spans from the milliwatt up to the watt level. Moreover, their tunability is another desirable feature. Unfortunately, on a time scale spanning from 1 s to 10 ms, QCLs linewidth is way wider in free-running operation (about 1 MHz) due to the 1/f noise contribution. There are three main approaches that can be used to overcome this limitation and to provide the desired absolute frequency reference:

- QCL emission can be stabilized and narrowed against a molecular absorption line, as reported for the first time at 8.5 μm, using a side of a Doppler-broadened $N_2O$ line [44] or using saturated absorption spectroscopy on $CO_2$ lines [45].
- The QCL can be stabilized onto an optical frequency comb (OFC) through a phase-locking chain [46–48].
- A high-Q whispering gallery mode resonator can be used to effectively narrow and reference the mid-IR QCL [49, 50].

Such schemes are described below in more detail.

### 3.1 Locking to molecular absorption profiles

A spectroscopic technique that is based on polarization-dependent signals was recently developed and proven to be extremely useful for narrowing QCLs emission and providing at the same time an absolute frequency reference [51]. It exploits the availability of a natural ruler, in the frequency domain, provided by the many strong ro-vibrational molecular absorption lines, whose center frequency can be absolutely measured with a sub-kHz precision [52]. Basing on this, it is possible to set up a simple system for high-sensitivity/precision spectroscopy for a specific molecular species, without using an OFC. A polarization-spectroscopy (PS) scheme produces, without any external modulation, the narrow dispersive sub-Doppler signal used to close the feedback loop on the QCL driving current for frequency stabilization. It has been demonstrated that, by locking the laser to a $CO_2$ line, the linewidth of a CW RT QCL can be narrowed below 1 kHz (FWHM) [51]. The laser used was an RT distributed-feedback (DFB) QCL emitting at 4.3 μm, provided by Hamamatsu photonics. It was operated at a temperature of 283 K and a current of 710 mA, delivering an output power of about 10 mW. A schematic of the experiment is shown in Figure 3. The QCL is mounted on a specific compact thermoelectrically cooled mounting. A low-noise home-made current driver was used. It ensures a current noise power spectral density always below 1 nA/√Hz while keeping a fast current modulation capability, thanks to a control circuitry placed in parallel to the QCL based on a field-effect transistor (FET).

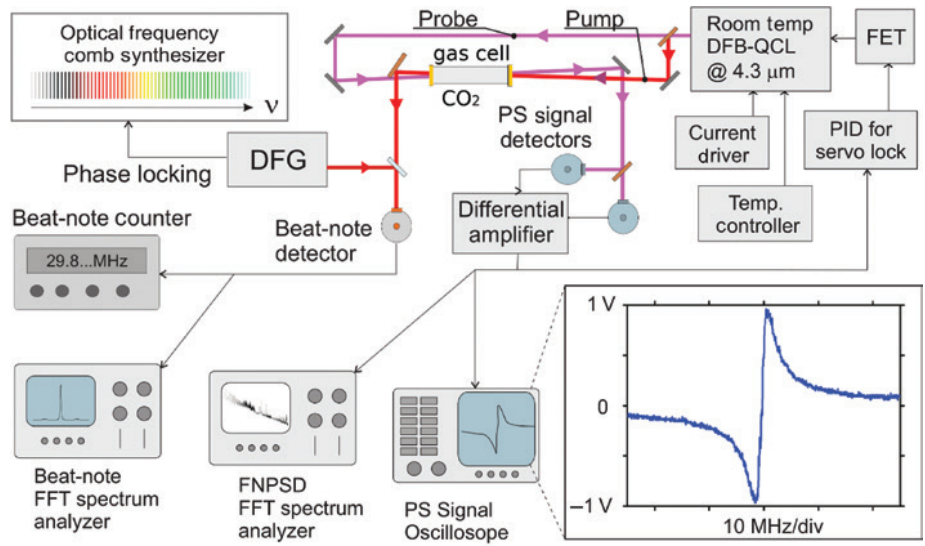

**Figure 3:** Polarization-locking setup.
The probe beam gives the signal used for the frequency locking. The pump beam is also used for the beat-note detection and the frequency counting. PS, polarization spectroscopy; DFG, comb-referenced single-frequency difference-frequency generation; FET, field-effect transistor; FFT, fast Fourier transform. Adapted with permission from [51], copyright 2012.

The chosen molecular transition is the P(29)e of the $(01^11 - 01^10)$ ro-vibrational band of $CO_2$ at 2311.5152 cm$^{-1}$. The inset in Figure 3 shows a typical scan of the PS signal at a pressure of 8.9 Pa, when the laser frequency is tuned across the molecular resonance. By carefully balancing the differential detection, a zero-offset signal is obtained. It ensures a linear conversion of the laser frequency fluctuations into amplitude variations in the region centered around the resonance frequency. For the QCL frequency stabilization, the PS signal is processed by a home-made PID controller and fed back to the FET gate for current control. From a preliminary analysis of the free-running frequency noise power spectral density (FNPSD) of a similar QCL [43], it is expected that a locking bandwidth of about 100 kHz is required for achieving a kHz-level linewidth. In order to ensure this condition, both the differential amplifier and the PID (proportional-derivative-integral) have been designed to have bandwidths larger than 1 MHz. However, there are two more fundamental aspects that can limit the loop bandwidth. The first is the roll-off of the QCL tuning rate with the modulation frequency [53]; the tuning rate is never flat, even at low frequencies, and shows a –3 dB cut-off at about 100 kHz. The second is the width of the linear region of the PS signal, which introduces a frequency roll-off starting from 300 kHz. Following the above considerations, the bandwidth of the frequency-locking loop is expected to be in the range of a few hundreds of kHz.

In order to characterize the frequency locking signal, two different measurements are carried out in parallel. The first one is the spectral analysis of the in-loop PS signal, the second one is the analysis of the beat note between the QCL and a narrow OFC-referenced DFG source providing a stable (10-Hz linewidth within 100 μs) and absolute reference. Each measurement was also performed with the QCL in free-running regime.

In Figure 4 – left, the FNPSD measurement results are shown. It is noteworthy to highlight the improvements in the free-running regime brought by the evolution of the current driver; using a new-generation low-noise driver provided by ppqSense, the FNPSD exhibits a clean 1/f trend, confirming that virtually no external noise is added. By closing the frequency-locked loop, the FNPSD is reduced in the spectral range below 250 kHz, which is then assumed to be the loop bandwidth, as expected. At about 450 kHz, the onset of a self-oscillation peak is evident. It can be well explained by the dephasing introduced by the approaching roll-offs mentioned above, and it is, at present, the factor limiting the loop performance. The FNPSD of the locked QCL is obtained by adding to the closed-loop error signal the detection noise floor. The latter is dominated, in the low-frequency range, by the residual intensity noise of the QCL and limits the actual frequency-noise reduction. The effect of the locking on the QCL emission line shape can be more intuitively described by the spectrum of the beat note between the QCL radiation and the DFG one. An acquisition is shown in Figure 4 – right. The 450-kHz servo bumps confirm the oscillation peak appearing in the FNPSD.

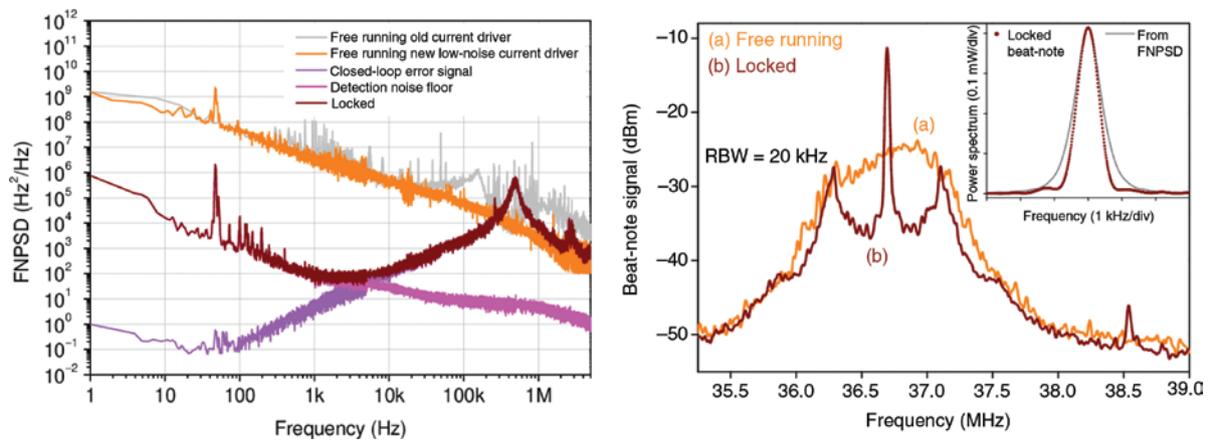

**Figure 4:** QCL FNPSDs (free-running and locked) and beat note with the DFG source.
Left: Comparison between the free-running QCL FNPSDs and the locked one. The locked QCL FNPSD is obtained by summing the spectrum of the closed-loop error signal to the PS detection noise floor, measured with empty gas cell. Right: FFT spectra of a 10-ms-long evolution of the beat note between the QCL, in free-running (trace a) and locked (trace b) conditions, and the narrow DFG source. Inset: zoomed view (linear scale) of the central peak observed over 1 ms with a resolution bandwidth of 721 Hz (dotted curve) and QCL power spectral profile retrieved from the locked FNPSD (straight line). Reprinted with permission from [51], copyright 2012.

By comparing the areas of the locked and free-running beat notes, we notice that 77% of the QCL radiation power is within the narrow peak centered on the molecular line. Switching from the free-running to the locked regime, the linewidth (FWHM) is reduced from about 500 kHz down to 760 Hz on a 1-ms time scale (inset). The inset also shows the comparison between the beat note and the locked QCL power spectral profile, retrieved from its FNPSD over a 1-ms time scale. For the latter, a 900-Hz FWHM is obtained, in good agreement with the beat-note linewidth. The beat-note frequency is also measured by a 1-s-gated frequency counter over about 2 h. The obtained Allan deviation [54] is 3 kHz at 1 s and decreases down to 0.9 kHz up to 320 s. Then, for longer times, it increases again, due to slow variations of the locking signal offset. This prevents our oscillator from achieving the stability performances of the best mid-IR standards [55]. The absolute frequency of the $CO_2$ line is measured by averaging a set of frequency counts performed by counting the beat note over several days and knowing the DFG frequency, thanks to the reference. The obtained value is $69,297,480.708 \pm 0.025$ MHz, with an uncertainty that takes into account both the repeatability of the offset zeroing and the OFC accuracy. This result is in agreement with the value given by HITRAN database [56] for this transition but with at least two orders of magnitude increased accuracy.

### 3.2 Locking to an OFC

Direct phase-locking of QCLs to OFCs is a valid alternative, compared to frequency locking to a molecular absorption line, allowing to enhance the frequency stability while preserving the full tunability of the laser source, at the cost of a more complex and bulky setup. Mid-IR metrological reference generation has been extensively demonstrated in a large variety of schemes by many groups. Mainly, we can report three approaches:

a. **Upconversion of mid-IR QCL emission to the visible or near-IR region**
   This approach is probably the most popular due to the large number and low-cost components and sources in the working spectral range. It has been widely demonstrated in the overall mid-IR region [57–62]. For example, in ref. [57], a QCL at 5.4 μm is upconverted to 1.2 μm by sum frequency generation (SFG) using an orientation-patterned GaAs pumped with a cw 1.5-μm fiber laser (Figure 5 top). An Er:fiber OFC is used to measure both the 1.2-μm and the 1.5-μm photons. This setup has been used in ref. [58] to perform proof-of-principle spectroscopy of $N_2O$ lines at 5.36 μm, as shown in Figure 5 – bottom.

   Coherent phase-locking of a QCL using a similar approach is reported also around 10 μm [61]. In this case, the stability of a $CO_2$ laser locked onto a saturated absorption line of $OsO_4$ (secondary frequency standard of this spectral region) is transferred to a QCL suitable for high-resolution spectroscopy on $NH_3$ and methyltrioxorhenium (MTO) for Boltzman constant determination and parity-violation test. The same group [62] realized a metrological setup referenced to a primary frequency reference disseminated by an optical fiber link of 43-km length (Figure 6). In this experiment, the QCL is phase locked by using a sum-frequency process in an $AgGaSe_2$ crystal to a near-IR OFC. Frequency standard dissemination via optical fiber, throughout Europe, has been recently demonstrated by several groups. Nowadays, it is well assessed that long-haul fiber-based optical frequency dissemination is a reliable tool for remote end users to perform high-precision measurements well beyond the ultimate capabilities of GPS referenced standards [63–66]. This complex metrological chain was used to perform saturated absorption spectroscopy using an FP cavity filled with $OsO_4$, achieving an accuracy on the center of the line of about 50 Hz. A detailed balance of systematic effects affecting line position was also reported (Table 1).

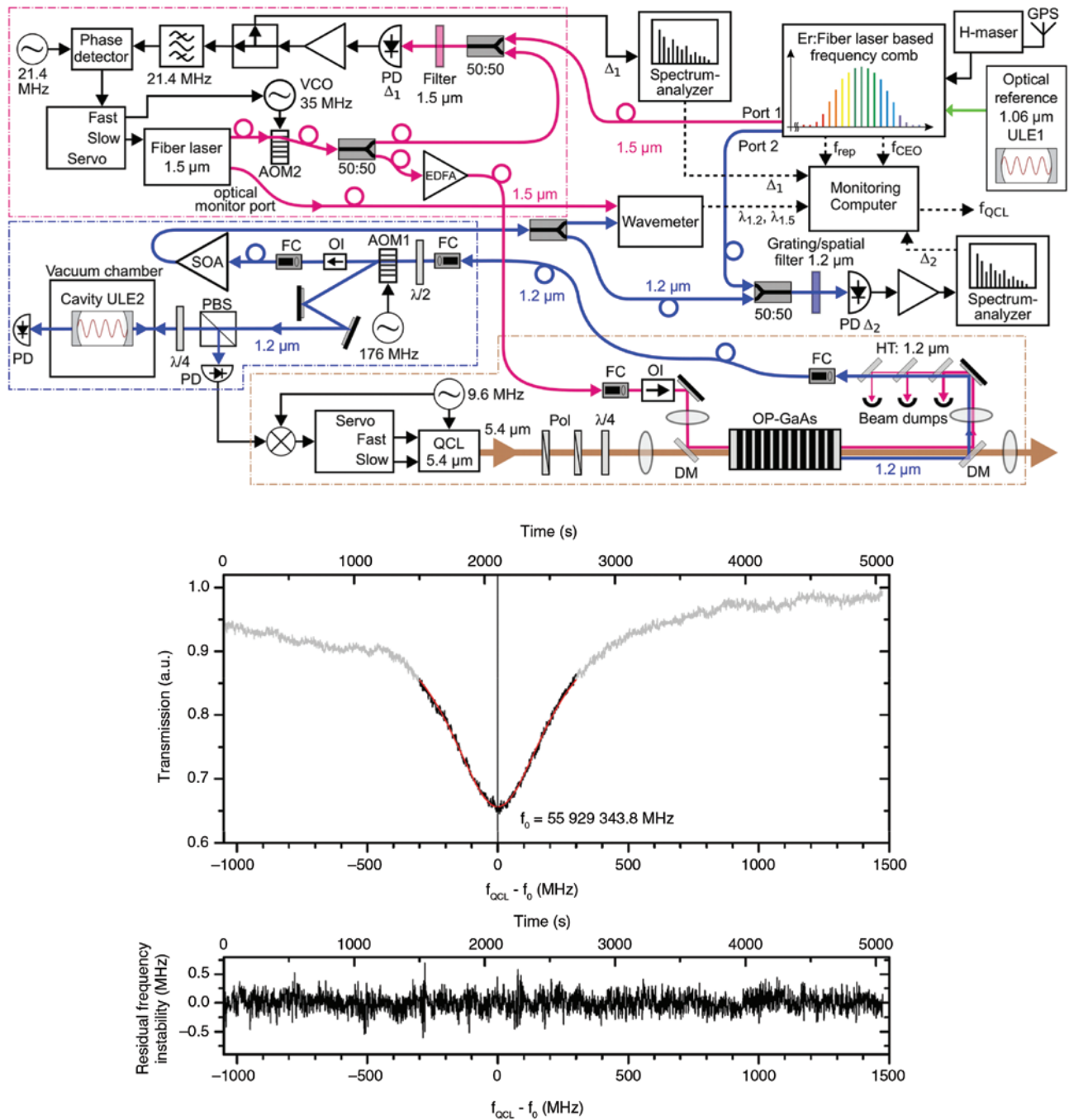

**Figure 5:** Example of an upconversionsetup and absorption profile of a n $N_2O$ transition.
Top: Hansen et al. [57] report an upconversion setup using a 5.4-μm emitting QCL. Pink, fiber laser wave at 1.5 μm; brown, QCL wave at 5.4 μm; blue, sum frequency wave at 1.2 μm. The spectrum analyzers are referenced to the H-maser. Reprinted with permission from [57], copyright 2015. Bottom: Absorption profile of $N_2O$ line around 5.36 μm. The central part of spectrum is fitted using a Voigt profile (red line). Bottom frequency residual instability during the frequency scan. Reprinted with permission from [58], copyright 2013.

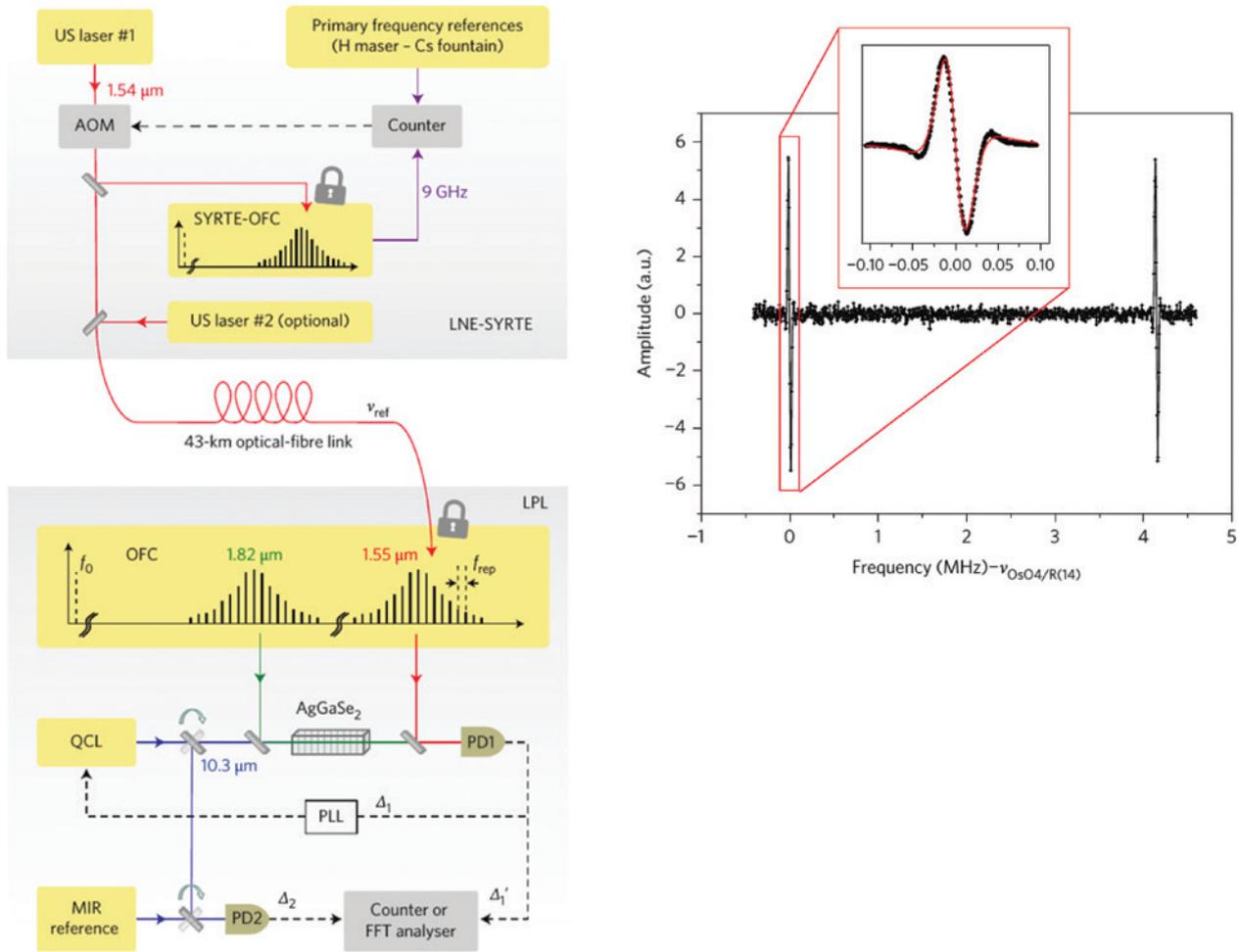

**Figure 6:** Spectroscopy on $OsO_4$ and locking chain.
Left: The Metrological chain described in [62]; a NIR metrological primary frequency reference is disseminated through a 43-km-long optical-fiber link. A 10-μm wavelength QCL is then phase-locked by performing sum-frequency generation in an $AgGaSe_2$ crystal. Right: Third derivative spectroscopy of $OsO_4$ is reported. Reprinted with permission from [62], copyright 2015.

**b. OFC-assisted mid-IR DFG**

In the study by Galli et al. [47], a QCL is directly phase-locked to a DFG mid-IR radiation obtained starting from two OFC-referenced near-IR sources [47]. This method provides simultaneously an absolute frequency reference and a residual phase noise independent of the OFC noise. Finally, a QCL linewidth narrowing below the OFC tooth one is obtained.

Indeed, linewidth below 1 kHz on a 1-ms time scale is observed from the analysis of the FNPSD. The QCL frequency stability and the absolute traceability were characterized, showing that both were limited by the Rb-GPS-disciplined 10-MHz quartz oscillator reference of the OFC. Precision and high-resolution spectroscopy performance of this QCL is tested by measuring the frequency of the saturation Lamb dip of a few $CO_2$ transitions with an uncertainty of $2\times10^{-11}$.

The laser is the one already mentioned in section 3.1, an RT DFB QCL emitting at 4.3-μm wavelength. It is operated at a temperature of 283 K and a current of 710 mA. The QCL was locked onto a reference radiation generated by DFG process in a periodically poled $LiNbO_3$ crystal [67, 68], by mixing an Yb-fiber-amplified Nd:YAG laser at 1064 nm and an external-cavity

**Table 1:** Absolute frequencies of five $OsO_4$ absorption lines in the vicinity of the R(14) $CO_2$ laser line.

| $OsO_4$ lines in the vicinity of the $CO_2$ R(14) laser line at 10.3 μm | Frequency shift from $\nu OsO_4$/R(14) calculated from refs. [39] and [41] (kHz) | Frequency shift from $\nu OsO_4$/R(14) measured in this work (kHz) |
|---|---|---|
| $^{190}OsO_4$ reference line (unassigned) | 0.000 (40) | −0.009 (22) |
| Unreported line | – | +4,147.399 (23) |
| $^{190}OsO_4$, R(46)$A_1^3$(−) | +101,726.83 (5) | +101,726.821 (32) |
| Unreported line | – | +123,467.401 (32) |
| Unreported line | – | +204,269.162 (33) |

Spectroscopic results on line shifts and center frequency of $OsO_4$ are reported. The frequencies are given with respect to the $OsO_4/CO_2$-R(14) reference line frequency, $\nu OsO_4/R(14) = 29.137,747,033,333$ THz, reported in ref. [39]. In the second column, we report the absolute frequencies calculated from refs. [39] and [41] with 1σ uncertainty. The third column displays the results of this work, where the uncertainty is the standard uncertainty of the mean. The R(46)$A_1^3$(−) line has previously been recorded at lower pressure [40]. Our measurement is thus expected to be pressure-shifted by approximately +10 Hz ref. [26].

diode laser (ECDL) emitting at 854 nm. This peculiar locking scheme, using a direct digital synthesis (DDS) architecture [69–71], provides an effective phase-locking of the ECDL to the Nd:YAG laser, while the OFC just acts as a transfer oscillator adding negligible phase noise to the DFG radiation. As a consequence, the mid-IR radiation is referenced to the Cs frequency standard through the OFC, but its linewidth is independent of the OFC one.

A schematic of the experimental setup is shown in Figure 7. A portion of the QCL beam, taken using a beam-splitter, is used for phase-locking. It is overlapped to the DFG beam through a second beam splitter and sent to a 200-MHz-bandwidth HgCdTe detector. A 100-MHz beat note is detected by using a few μW of both QCL and DFG sources. The beat note is processed by a home-made phase-detection electronics, which compares it with a 100-MHz local oscillator (LO) and provides the error signal for closing the phase-locked loop (PLL). A home-made PID (proportional-integral-derivative) electronics processes the error signal and sends it to the gate of a FET for a fast control of the QCL driving current.

In Figure 8 – left, the beat note acquired using an FFT spectrum analyzer is shown. The width of the carrier frequency is limited by the instrumental resolution bandwidth, as expected from a beat note between two phase-locked sources. The locking bandwidth is limited by the dependence of the QCL tuning rate on the modulation frequency. A 250-kHz locking bandwidth is achieved, as confirmed by the servo bumps in the beat note. The phase-locking performance in terms of residual RMS phase error is measured by using the fractional power η contained in the coherent part of the beat-note signal, i.e. in the carrier. By evaluating the ratio between the area under the central peak of the beat note and the area under the whole beat-note spectrum (1.5-MHz wide), a phase-locking efficiency of η = 73% is obtained, yielding a residual RMS phase noise of 0.56 rad. The main portion of the QCL radiation is used for frequency-noise characterization and for spectroscopy. For the first purpose, the QCL beam is coupled to a high-finesse cavity, which works as frequency-to-amplitude converter, when its length is tuned, in order to have a transmission corresponding to half the peak value. The cavity free spectral range is 150 MHz, and its finesse is about 9000 at λ = 4.3 μm, as measured with the cavity-ring-down technique, leading to a mode FWHM of 18.8 kHz. The cavity output beam is detected by a second HgCdTe detector, and the resulting signal is processed by a FFT spectrum analyzer.

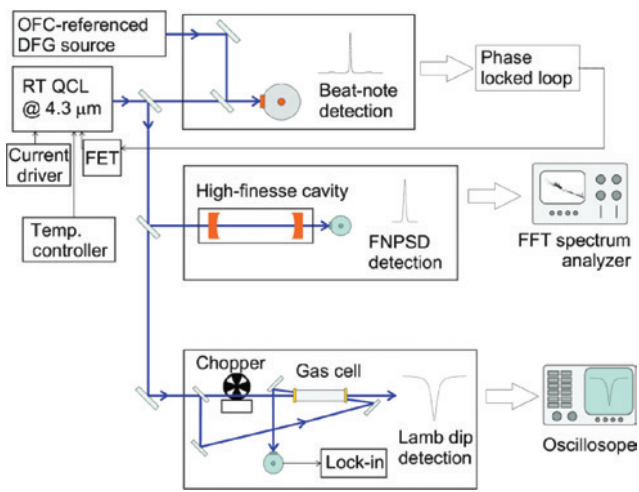

**Figure 7:** Schematic of the experimental setup.
There are three main parts: the beat-note detection between QCL and DFG for the phase-locking, the high-finesse cavity for FNPSD analysis, and the saturation spectroscopy signal detection for the absolute frequency measurement of the $CO_2$ transitions. Reprinted with permission from [47], copyright 2013.

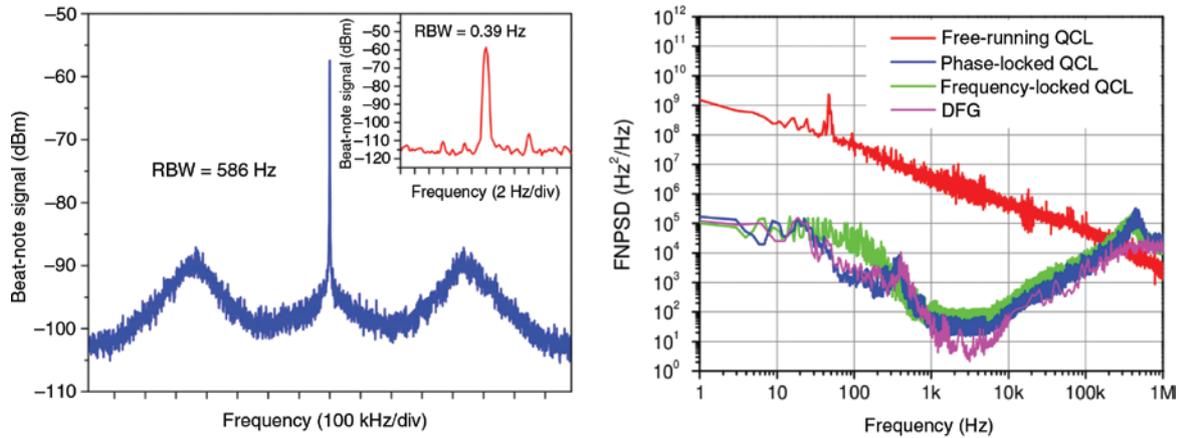

**Figure 8:** QCL beat note with the DFG source and FNPSDs (free-running and locked).
Left: Beat-note signal between the DFG radiation and the phase-locked QCL. The inset shows the same beat note with a narrower span and resolution bandwidth. In both cases, the width of the peak (FWHM) is limited by the resolution bandwidth of the spectrum analyzer. Right: QCL FNPSDs in free-running and phase-locked conditions, acquired by using a $CO_2$ line and the high-finesse cavity as frequency-to-amplitude converters, respectively. Reprinted with permission from [47], copyright 2013.

In Figure 8 – right, the FNPSD of the phase-locked QCL, acquired by using the high-finesse cavity, is shown. The same cavity was also used to measure the DFG FNPSD and the QCL FNPSD when frequency-locked to a molecular absorption line. Such an independent converter enables a fair comparison between the two different locking techniques. The plotted FNPSDs are compensated for the high-frequency cavity cut-off, due to the photon cavity ring-down rate ($f_c = 9.4$ kHz). The free-running QCL FNPSD, recorded by using the slope of the Doppler broadened $CO_2$ absorption line as converter, is shown.

The comparison between free-running and phase-locked conditions confirms a locking bandwidth of 250 kHz, with a frequency noise reduction of about four orders of magnitude for frequencies up to 10 kHz. Moreover, the phase-locked-QCL FNPSD perfectly overlaps the DFG one, with only some excess noise above 200 kHz. If we compare the QCL FNPSD when phase/frequency-locked to the DFG/molecular transition, they are almost coincident for Fourier frequencies above 1 kHz and up to 450 kHz, where the self-oscillation of both control loops is observed. This confirms that the locking bandwidth is limited by the laser modulation bandwidth. Nevertheless, a QCL linewidth narrower than 1 kHz (FWHM) on a time scale of 1 ms is retrieved in both cases by integrating the FNPSDs for frequencies above 1 kHz. As a consequence, we note that phase-locking the QCL does not improve laser narrowing with respect to frequency-locking. On the other hand, between 30 Hz and 1 kHz, the two curves show different trends; in this range, the phase-locked QCL FNPSD lies below that of the frequency-locked one, except for an evident noise peak centered at 400 Hz, which is also present in the DFG source. Reasonably, the phase-locked QCL FNPSD would be flat at 300–400 Hz, employing a master with a flat FNPSD in that frequency interval. Apart from this peak, the comparison in this frequency range confirms a better control of the frequency jitter for the phase-locked QCL, overcoming the limits of the frequency-locked QCL set by the presence of a residual amplitude noise. For Fourier frequencies below 30 Hz, the high-finesse cavity is no longer a good frequency-to-amplitude converter, as it saturates. In this frequency range, the phase-locked QCL can be reasonably considered more stable than the frequency-locked one, as the first inherit the stability from the cesium-clock-referenced OFC used as reference.

A sub-Doppler saturated-absorption scheme using a phase-locked QCL was used to measure several lines of the $CO_2$ molecule around 4.3 μm in a 12-cm-long, single-pass cell [47, 67–72]. Within the tuning range of the QCL (10–25°C for temperature and 700–900 mA for current), the absolute center frequency of six $CO_2$ transitions belonging to the P branch of its ($01^11 - 01^10$) ro-vibrational band could be measured. In the saturated-absorption scheme, the Lamb dip at the center of the Doppler-broadened molecular line is detected. Thanks to the high precision/accuracy achieved by this setup, a set of acquisitions could be performed by varying the pressure of the $CO_2$ gas in a very small range (1–26 Pa – Figure 9). A weak linear dependence of the line centers on pressure is observed, as it is generally expected for any molecular line (pressure shift). The extrapolated values at P = 0 yield the absolute frequencies for all the measured transitions, corrected for systematic self-pressure shifts.

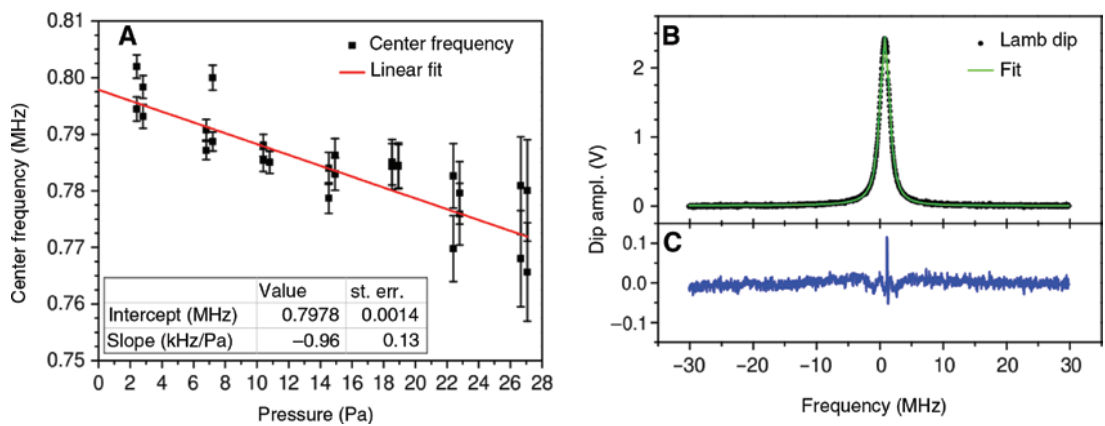

**Figure 9:** High-resolution spectroscopy on $CO_2$.
(A) Dependence of the center of the Lamb dip of the $CO_2$ ($01^11 - 01^10$) P(29)e transition on pressure, with the corresponding linear fit. For clarity, the constant value of 69297478 MHz has been subtracted from the absolute frequency values. (B) Example of a single Lamb dip acquisition. Experimental conditions: lock-in amplifier time constant of 10 ms, chopper frequency of 616 Hz, frequency scan with 60-MHz span and 50-kHz steps. The fit to a Voigt function (green line) and the residuals (C) are also plotted. Adapted from [47].

In Table 2, the measured line-center frequencies (obs. freq.) and the measured self-pressure shift coefficients (press. shift) are listed. The deviations with respect to two different sets of calculated values (obs.-calc.) are also reported. From the first comparison, we note that the HITRAN frequency values are systematically blue-shifted by 4–5 MHz with respect to the observed ones, probably due to a miscalibrated spectroscopic apparatus [56]. Those measurements are three to four orders of magnitude more precise than the values extracted from the HITRAN database, and thus, they could be used to improve the molecular parameters of the ($01^11 - 01^10$) ro-vibrational band (especially the band center).

### c. Generation of a mid-IR frequency comb

An OFC can be translated from the visible/near0infrared domain to the mid-IR region, taking advantage of DFG. The challenge, in this case, is to overcome the lack of optical components in this spectral region. Considering the wide ro-vibrational bands characterizing mid-IR molecular spectra, it is clearly interesting and useful to have OFCs operating directly in this spectral region. IR-downconverted OFCs can serve as direct references for single-frequency mid-IR lasers, such as DFB QCLs, or they can be used directly for mid-IR spectroscopy. An OFC radiation source can provide more information at a fixed integration time than a single-frequency one, thanks to its wide, instantaneous spectral coverage. In order to perform high-sensitivity and high-resolution spectroscopy, it is again crucial to have intense and narrow (low-frequency-noise) radiation. Moreover, if also a high accuracy is required, an absolute reference is needed.

**Table 2:** Observed line-center frequencies and self-pressure shift coefficients for six transitions belonging to the ($01^11 - 01^10$) ro-vibrational band of $^{12}C^{16}O_2$.

| Transition | Obs. freq. (MHz) | Press. shift (kHz/Pa) | Obs.-calc.[a] (MHz) | Obs.-calc.[b] (MHz) |
|---|---|---|---|---|
| P(29)e | 69 297 478.7998 (31) | −1.37 (52) | −4.0611 | 0.1362 |
| P(30)f | 69 267 227.7792 (36) | −1.04 (46) | −4.7036 | −0.0321 |
| P(31)e | 69 239 948.5044 (11) | −0.22 (9) | −4.0039 | 0.3028 |
| P(32)f | 69 209 198.3778 (8) | −0.16 (9) | −4.6879 | −0.1170 |
| P(33)e | 69 181 700.3740 (18) | 0.05 (11) | −3.9287 | 0.0376 |
| P(34)f | 69 150 447.0863 (14) | −0.22 (10) | −4.6021 | −0.1919 |

[a]Comparison with the HITRAN database. The uncertainty reported by the original database for each transition is in the 3–30-MHz interval.
[b]Comparison with ref. [73], where the frequencies of the same transitions measured with a low-resolution (162 MHz) Fourier-transform interferometer are reported.
Comparisons with calculated values are also reported [72].

Pulsed mode-locked lasers have not been developed in the mid-IR yet, but their near-IR spectra can be transferred to the mid-IR (MIR combs) taking advantage of non-linear frequency mixing.

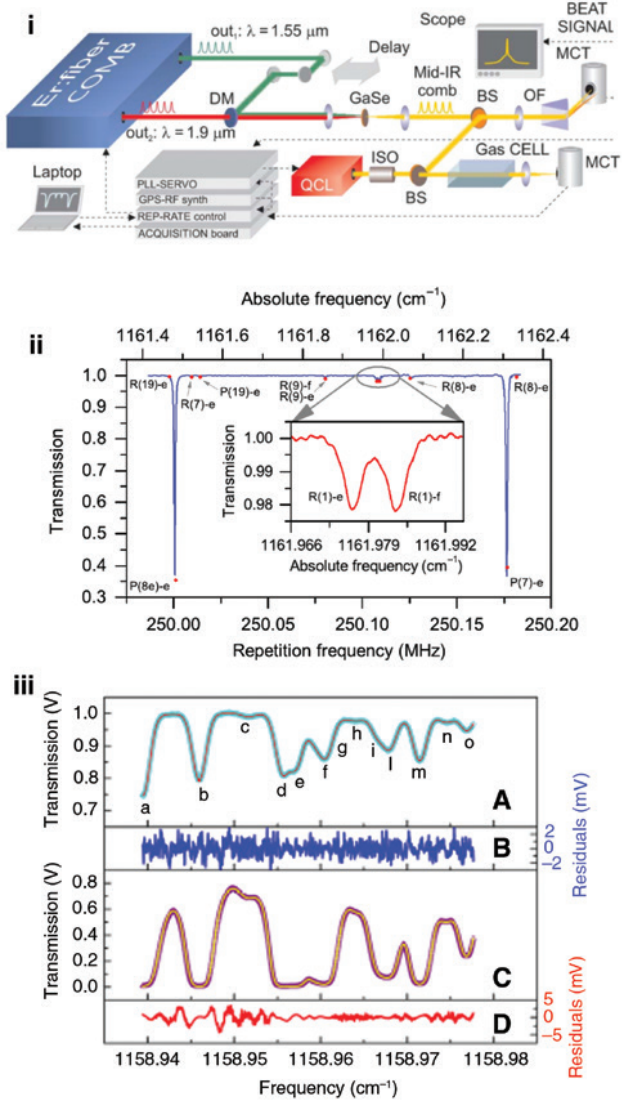

**Figure 10:** (i) Experimental setup for absolute $N_2O$ spectroscopy at 8.6 μm. A mid-IR comb generated by a non-linear process in GaSe starting from the two output of a Near-IR comb is used as metrological reference for spectroscopy of $N_2O$ e $CH_3F$ in the region of 8.6 μm. Reprinted with permission from [75], copyright 2015. (ii) $N_2O$ transmission spectrum at 8 mbar (800 Pa) pressure, resulting from a wide scan of the comb rep-rate (bottom scale) and corresponding QCL absolute frequency (top scale). Red dots: HITRAN calculated absorption-peaks in correspondence of our experimental conditions. Inset: detail of the R(1)-e-R(1)-f doublet with a 2% peak fractional absorption. Reprinted with permission from [75], copyright 2015. (iii) (A) $CHF_3$ transmission profile around 8.6 μm recorded at a temperature of 296 K and a pressure of 8.8 Pa (green curve) together with a multi-line (13 line) Voigt fitting (red curve). (B) Fit residuals. (C) and (D) are the same as (A) and (B) but with a pressure of 200 Pa. Reprinted with permission from [74], copyright 2015.

In literature, a number of schemes and non-linear crystals used for mid-IR comb generation ($AgGaSe_2$, GaSe, Op-GaAs, $LiNbO_3$) can be found [48, 74–82]. In ref. [74–76], the authors describe a mid-IR comb (8–14 μm), based on DFG in a GaSe nonlinear crystal pumped by a dual output of a Er:fiber laser oscillator (Figure 10-i). The first output is at 1.55 μm, while the second output is generated by using a nonlinear Raman fiber, with a spectrum in the 1.76–1.93-μm range. These two pulses, overlapped in time, are used in a DFG process in a GaSe crystal. Spectroscopy on $N_2O$ and $CH_3F$ was performed by using an 8.6-μm QCL (Figure 10-ii–iii and Table 3) phase-locked to this mid-IR OFC [74–76].

A similar approach, using a Ti:sapphire laser cavity and a difference-frequency process, was realized by Galli et al. [80] to generate a frequency comb around 4330 nm with a power-per-tooth of 1 μW. In this case, a phase-locking scheme based on direct digital synthesis (DDS) was used, providing a tooth linewidth of about 2.0 kHz on a timescale of 1 s (Figure 11).

A promising alternative is represented by QCLs (QCL-combs) [83]. Several experiments have already proven the intrinsic coherence of the emission of QCL combs [84, 85], but in order to be used for high-resolution spectroscopy applications, a proper stabilization is required to overcome technical noise. Indeed, for metrological purposes, a fine control of the main optical parameters is required [86]. In ref. [87], a DFG comb was used to investigate the actual features of the multimodal QCL emission in a dual-comb-like setup. This characterization is essential for metrological applications of QCL combs. QCL combs have already been applied for dual comb spectroscopy [84], but their potential as metrological mid-IR references and sources has still to be demonstrated.

In this direction, two experiments have been recently carried out in order to retrieve the phase coherence level (coherence among the modes) of such sources [88, 89]. When operating in frequency-comb regime, QCL combs present a remarkable level of phase coherence between the emitted longitudinal modes, confirming the frequency comb nature of such sources.

### 3.3 Stabilization with crystalline whispering gallery mode resonators

An alternative and effective tool for laser stabilization and linewidth narrowing is represented by high-Q whispering gallery mode resonators (WGMR). WGMRs made of crystalline materials have started to be used for mid-IR

**Table 3:** Main spectroscopic parameters of $CHF_3$ lines at 8.6 μm.

| Line | Centerwavenumber (cm$^{-1}$) | Intensity ($10^{-21}$ cm/molecule) | Pressure shift (kHz/Pa) | Pressure broadening (MHz/Pa) |
| --- | --- | --- | --- | --- |
| a | 1158.939443 (4) | 15.0 (1) | −15 (1) | 0.220 (3) |
| b | 1158.9459608 (2) | 11.8 (1) | 3.1 (1) | 0.115 (3) |
| c | 1158.951501 (6) | 0.4 (1) | −28 (35) | 0.236 (11) |
| d | 1158.955564 (6) | 8.0 (2) | −13 (2) | 0.070 (4) |
| e | 1158.957184 (4) | 9.4 (1) | −11 (1) | 0.090 (8) |
| f | 1158.95930 (14) | 5.2 (5) | 18 (7) | 0.196 (21) |
| g | 1158.960623 (8) | 5.3 (3) | 12 (3) | 0.081 (8) |
| h | 1158.963927 (9) | 0.4 (1) | 39 (5) | N.A. |
| i | 1158.966679 (7) | 3.1 (1) | 32 (24) | 0.103 (6) |
| l | 1158.968035 (5) | 3.6 (1) | 16 (13) | 0.113 (15) |
| m | 1158.9714499 (2) | 7.1 (1) | 1.0 (1) | 0.093 (7) |
| n | 1158.974553 (4) | 0.4 (1) | 900 (600) | 0.3 (2) |
| o | 1158.976854 (4) | 2.0 (1) | −13 (15) | 0.089 (12) |

Statistical uncertainties, quoted in parentheses, correspond to one standard deviation. Reprinted with permission from [74], copyright 2015.

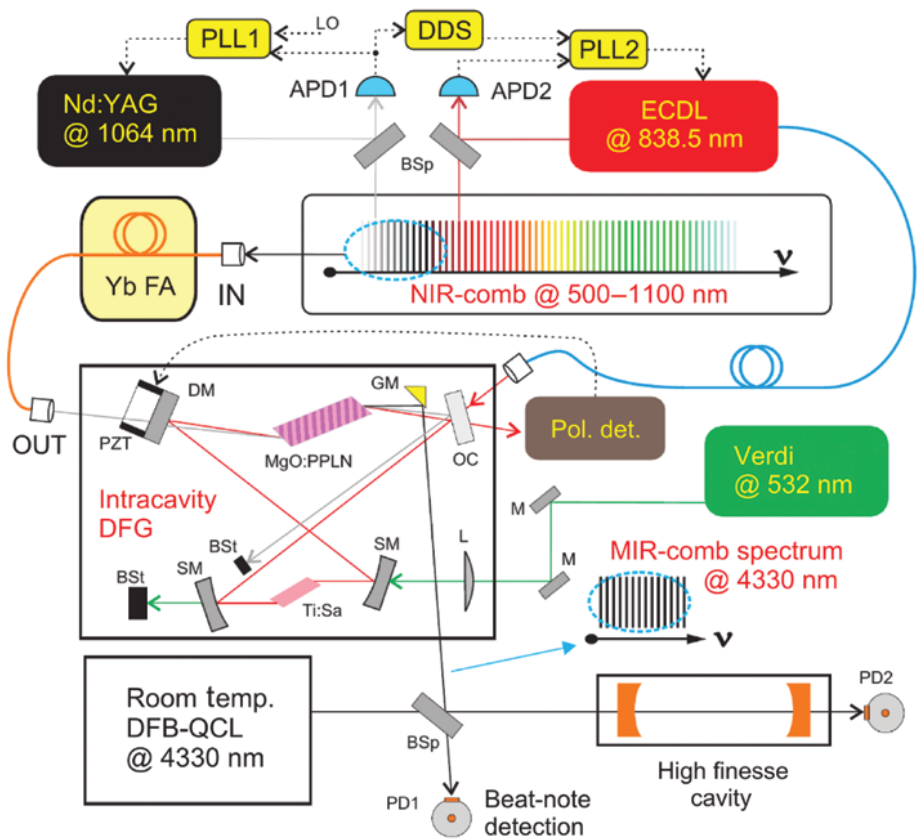

**Figure 11:** Mid-IR OFC generation setup.
It is based on a mode-locked Ti:Sa laser with a repetition rate of about 1 GHz, spectrally broadened (500–1100 nm) by a photonic-crystal fiber. The oscillator is referenced to an Rb/GPS-disciplined 10-MHz quartz clock with a stability of $6 \times 10^{-13}$ at 1 s and an accuracy of $2 \times 10^{-12}$. The portion of the NIR-comb spectrum above 1-μm wavelength is amplified by an $Yb^{3+}$ fiber amplifier. This light is injected into the Ti:Sa laser cavity and the output, together with a tunable diode at about 849 nm, is used in a DFG process in a MgO:PPLN crystal. The result is a MIR-comb centered around 4330 nm. Considering the 838–863-nm tuning range of the pump laser injecting source, this scheme provided a comb tunable from 4.2 to 5.0 μm. Reprinted with permission from [80], copyright 2013.

applications in the last 2 years. They are particularly interesting because of their potential to achieve high optical quality factors ($Q \simeq 10^{11}$ in the near-IR [90]) as well as to cover a wide transparency range, from the UV to the mid-IR. The ultimate Q factor is determined by the intrinsic material loss and scattering. The very sharp frequency response of resonant modes makes WGMRs appealing for sensing applications. Moreover, thanks to their narrow mode widths, crystalline WGMRs are attractive for frequency reference applications.

In the work described below, complementary methods for stabilizing a QCL emitting at 4.3-μm wavelength, including all-electronic locking onto the transmission and reflection modes of the resonator, are shown [49]. A $CaF_2$ toroidal WGMR (from OEwaves) was used. The resonator had a diameter of 3.6 mm, corresponding to a free-spectral range (FSR) of 18.9 GHz at the experimental wavelength, and was mounted inside a custom-made housing in order to reduce both thermal and mechanical fluctuations and in order to protect it from dust and humidity. The QCL was free-beam coupled to the resonator through a coupling prism, placed close to the resonator surface. By acting on temperature, it was possible to tune both the mode width and the resonance frequency in order to select the best coupling condition. Optimal coupling required a beam waist of about 10 μm (radius at $1/e^2$ of the total beam power). In operating conditions, the measured WGMR transmission mode was 3.1 MHz FWHM, corresponding to $Q \simeq 2.2 \times 10^7$. The measured value for the Q-factor is in agreement with other measurements made on similar WGMRs at the same working wavelengths [91, 92].

The improvement in terms of frequency stability and linewidth was quantified by measuring the laser FNPSD making use of a frequency-to-amplitude converter. In this setup, the converter is the side of a strong $CO_2$ absorption line, the (000–001) P(42) transition occurring at 2311.105 cm$^{-1}$, with a line strength of $4.75 \times 10^{-19}$ cm (HITRAN units). The $CO_2$ pressure inside the cell was chosen to maximize the slope of the absorption line (P $\simeq$ 1 mbar) for an optimal frequency-to-amplitude conversion. The measured laser FNPSD is shown in Figure 12 – left. The locking bandwidth exceeds 100 kHz, and the loop is able to pull down the laser FNPSD by more than three orders of magnitude with respect to the free-running laser (black trace). A linewidth around 700 kHz was inferred for the free-running laser (1-s timescale), which is reduced to 15 kHz in the locked regime (10 kHz for 1-ms timescale) [49]. This stability at long timescales marks the difference with respect to previous results on QCLs locked to mid-IR cavities [93], which suffer from a larger sensitivity to external acoustical and mechanical noise. It is interesting to note that the achieved noise reduction is very similar for both electronic and optical locking, allowing to choose among the two techniques according to the actual experimental needs.

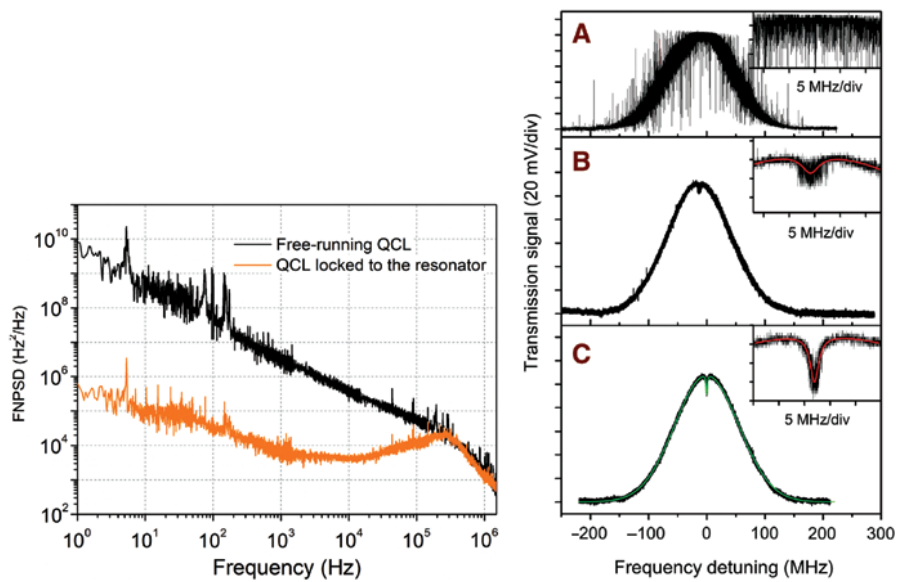

**Figure 12:** QCL FNPSDs (free-running and locked) and WGMR-assisted spectroscopy.
Left: Frequency-noise power spectral density for a free-running (black trace) and locked (orange trace) QCL. The frequency cut-off around 300 kHz is due to the limited bandwidth of the detector. Reprinted with permission from [49], copyright 2016. Right: The whispering Gallery mode resonator of [49] is used to assist sub-Doppler spectroscopy of $CO_2$ with a pump-probe scheme [50]. The three cases corresponding to the different traces are free-running laser operated with the commercial driver (A); laser operated with the home-made driver in free-running (B) and locking (C) conditions. All three graphs have the same horizontal scale, representing the detuning from the center frequency of the transition. In each inset, a zoom of the top line profile is shown, over the same horizontal and vertical scales, with the red curves showing the Lorentzian fit of the dip (B, C). The thin green trace superimposed on the data in (C) is a simulation of the sub-Doppler feature based on the HITRAN data. Reprinted with permission from [50], copyright 2016.

A test of the suitability of the QCL-WGMR system for high-resolution spectroscopy was performed [50]. In this test, a standard pump-probe setup for sub-Doppler spectroscopy was realized. In locking conditions, the QCL was tuned onto the same strong $CO_2$ transition mentioned above. A Lamb dip with a 2-MHz FWHM was recorded (Figure 12 – right), where the main width contribution was due to residual Doppler broadening. An uncertainty of 9 kHz on the transition center frequency was obtained, corresponding to a relative precision of about $10^{-10}$ over a few seconds acquisition time.

# 4 THz QCLs

## 4.1 Far-IR domain

The so-called terahertz spectrum, also known as far-IR, conventionally spans the frequency range from 0.1 to 10 THz, corresponding to the wavelength interval from 30- to 3000-μm wavelength. Historically, this region is known as "terahertz gap", due to the lack of sufficiently strong and compact sources and sensitive detectors [17]. One of the oldest applications of terahertz radiation is spectroscopy. Many chemical species have indeed very strong characteristic rotational and vibrational absorption lines in the THz range, whose absorption strengths are $10^3$–$10^6$ stronger than in the microwave region. As THz transitions represent a useful molecular "signature", astronomy and space science have recently moved to THz technology [94]. As a topical example, one-half of the total luminosity of the galaxy and 98% of the photons emitted as the Big Bang fall into the terahertz gap [95].

Much of this radiation is emitted by cool interstellar dust inside our and other galaxies, and thus the study of the discrete lines emitted by light molecular species can give an important insight into star formation and decay, despite the clear need of satellite platforms or high altitudes, due to the strong atmospheric absorption resulting from pressure broadened water and oxygen lines. Furthermore, terahertz thermal emission from gases in the stratosphere and upper troposphere such as water, oxygen, chlorine, and nitrogen compounds is useful for the study of chemical processes related to ozone depletion, pollution monitoring, and global warming [96]. Other spectroscopic applications include plasma fusion diagnostics [97] or identification of different crystalline polymorphic states of a drug.

The lack of coherent sources in this range was first filled by optically pumped fixed-frequency FIR lasers at the basis of, e.g. laser magnetic resonance (LMR) spectrometers, having a wider tunability but only working on paramagnetic species (see e.g. ref. [98]). Generation of microwave sidebands on the strongest FIR laser lines in Schottky diodes could produce tunable FIR radiation up to about 3 THz (100 cm$^{-1}$) [99, 100]. Continuous spectral coverage from 300 GHz up to about 9 THz was achieved by different configurations based on nonlinear mixing of microwaves with infrared radiation from carbon-dioxide lasers in metal-insulator-metal (MIM) diodes [101–103]. The unique combination of very wide tunability, few tens of kHz frequency uncertainty, and kHz level linewidth produced plenty of accurate frequency measurements of atomic and molecular transitions (see e.g. refs. [104–108]). A pioneering "hybrid" approach, generating far-IR radiation by mixing onto a MIM diode a frequency-locked $CO_2$ laser and a QCL emitting around 8-μm wavelength, allowed tunable spectroscopy of rotational lines of hydrogen bromide [109]. More recently, plenty of work has been done in order to develop high performance DFG setups [110–116]. However, the aforementioned setups were generally bulky and unreliable and emitted low output powers (in the nW to μW range).

## 4.2 Single-frequency THz QCLs

The availability of a new generation of compact, reliable THz sources with large output power represented the key for the development of the THz range. As discussed in the following, THz emitting QCLs are proving to be good candidates to fill this gap. The first report on a THz QCL [31] exploited a careful design of the active region based on a chirped superlattice and an asymmetric low-loss waveguide, achieving emission at 4.4 THz. In this first experiment, the mini-band width was kept lower than the optical phonon energy in order to avoid photon re-absorption. Since then, despite the cryogenic operation temperatures (199 K) [117], THz QCLs have attracted considerable attention thanks to the high output power (>100 mW), spectral purity, stability, compactness, and reliability and have now a realistic chance to boost technological applications. In fact, frequency- and phase-stabilized, high-power, and reliable solid-state terahertz sources can indeed find application in a large number of fields, from far-infrared astronomy [118] and high-precision molecular gas spectroscopy [119] to high-resolution coherent imaging and telecommunications [120, 121].

In addressing such application requirements, high-frequency stability sources are almost mandatory. In this context, a thorough knowledge of the intrinsic linewidth, ultimately limited by quantum noise, is the key, determining the achievable spectral resolution and coherence length. Environmental effects such as temperature, bias-current fluctuations, and mechanical oscillations are widely known to have a significant effect on emission linewidths in QCLs. This means that any experimental free running linewidth measurement is dominated by extrinsic noise [122–124]. Up to 2012, only a few experimental studies had indeed been reported on frequency narrowing and spectral purity of terahertz QCLs, providing upper limits of 30 kHz, 20 kHz, and 6.3 kHz for the "instantaneous" linewidth, respectively [122, 125, 126]. Environmental effects can be minimized by using frequency-stabilization or phase-locking techniques, resulting in narrower linewidths that are limited by the loop bandwidth of the specific experimental system [123].

### 4.3 Intrinsic linewidth

Recently, the spectral purity of a THz QCL was investigated via the measurement of its FNPSD, providing an experimental evaluation and a theoretical assessment of its intrinsic linewidth [127]. Intensity measurements were performed to retrieve information in the frequency domain by converting the laser frequency fluctuations into detectable intensity (amplitude) variations. As a discriminator, the side of a Doppler-broadened methanol molecular transition was used. Specifically, the ro-vibrational molecular transition line of $CH_3OH$ centered at 2.5227816 THz was used as a discriminator. Given the intrinsic low-noise nature of the measurement, the converter (or discriminator) must introduce negligible noise providing, at the same time, a gain factor suitable for good detection. A schematic diagram of the experimental setup used is shown in Figure 13.

The collimated THz QCL beam is sent to the gas cell for spectroscopy experiments. It is then split by a wire grid polarizer; the reflected beam is chopped and sent to a pyroelectric detector for the acquisition of the line profile and for frequency stabilization; the transmitted beam is acquired by means of two detectors (a silicon bolometer and a hot-electron bolometer (HEB), depending on the required bandwidth) and used for the frequency-noise measurement. It is worth noting that the gas cell window was properly tilted with the specific purpose to avoid any optical feedback effect on the measured frequency noise.

During frequency-noise measurements, QCL frequency needs to be locked at the half-height position of the absorption line in order to keep the conversion factor constant at its maximum value. The latter procedure was done by implementing a software PI loop on the QCL current and by using the line around the locking point as a feedback signal. This allowed for efficient stabilization of the mean QCL frequency at the right point, without affecting the QCL frequency noise above 10 Hz.

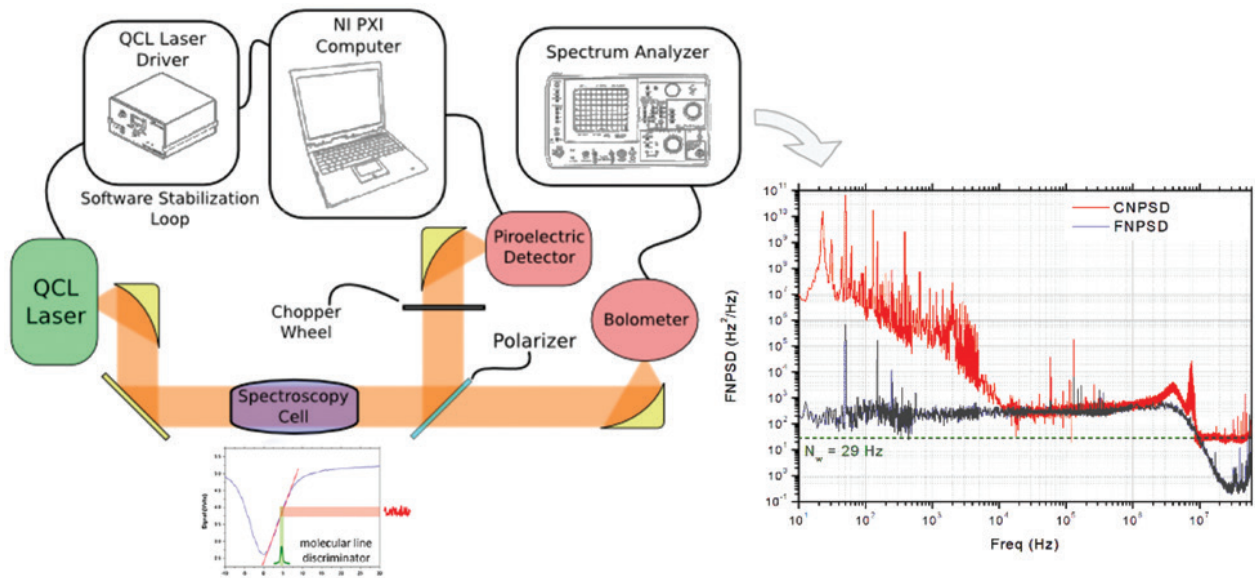

**Figure 13:** FNPSD of a THz QCL.
Left: Schematic diagram of the experimental setup, using a molecular transition as frequency-to-amplitude discriminator. Right: experimental FNPSD of the THz QCL (red trace), compared with the contribution to frequency noise of the CNPSD of the current driver (black trace). The dashed green line marks the white noise level. Reprinted with permission from [127], copyright 2012.

The combination of the large detection bandwidth of the HEB with the low-noise fast-Fourier-transform acquisition enabled spectral measurements spanning over seven frequency decades (from 10 Hz to 100 MHz) and 10 amplitude decades. A measurement of the residual amplitude noise was also performed by shifting the QCL frequency out of the discriminator side. The latter was then subtracted from the former in order to retrieve the correct FNPSD. The full spectrum of the QCL was obtained by sticking together several acquisitions taken in smaller spectral windows in order to ensure a high overall resolution. The resulting FNPSD spectrum is plotted in Figure 13 (inset), together with the current-noise power spectral density (CNPSD) of the current driver, converted to the same units by using the current tuning coefficient. Residual external noise gives rise to the sharp peaks visible throughout the trace. Three distinct domains can be clearly recognized in (i) the f = 10-Hz to 10-kHz range, where the FNPSD is dominated by noise not arising from the current driver and therefore ascribed to the QCL itself. The excess frequency noise with respect to the CNPSD level, which is absent at larger frequencies, can be attributed to spurious low-frequency background radiation signals and/or electronic noise, together with electric field fluctuations due to gain Stark shift and to cavity mode pulling; (ii) the 10-kHz to 5-MHz range where the FNPSD is fully dominated by the contribution of the current driver; (iii) above 8 MHz, where an asymptotic flattening is observed in the FNPSD, with a significant deviation from CNPSD, thus suggesting a flattening to a white noise level, therefore leading to an intrinsic linewidth of 90 ± 30 Hz.

Later the same year, similar results were confirmed by an independent study conducted using a near-IR frequency Comb [128]. The technique is based on heterodyning the laser emission frequency with a harmonic of the repetition rate of a near-IR laser comb. This generates a beat note in the radio frequency range that is demodulated using a tracking oscillator, allowing the measurement of the frequency noise. An intrinsic linewidth of ~230 Hz for an output power of 2 mW was retrieved. Both these results further qualify THz QCLs as ideal metrological grade sources.

## 4.4 Metrological grade THz QCLs

The development of a metrological referencing technique is very promising in order to fully exploit these devices potentialities. The first attempts of performing molecular spectroscopy with an absolute frequency scale, using a THz QCL, were performed by Hübers and co-workers in 2006, using a 2.5-THz distributed feedback device, frequency locked to a molecular laser transition [129]. The experimental setup thereby adopted is reported in Figure 14A; a high-power $CO_2$ laser optically pumps a molecular far-infrared laser emitting at 2.5227816 THz, while its emitted radiation and the radiation of the CW QCL are superimposed by means of a wire-grid polarizer. The optical mixing is obtained on a Schottky diode, allowing for QCL frequency referencing. The QCL frequency was scanned through a low-pressure methanol absorbing gas, and one acquisition example is shown in Figure 14B. The setup was able to measure the transition self-pressure shift, and the final accuracy achieved on the transition center was of 1 MHz ($4 \times 10^{-7}$ relative accuracy).

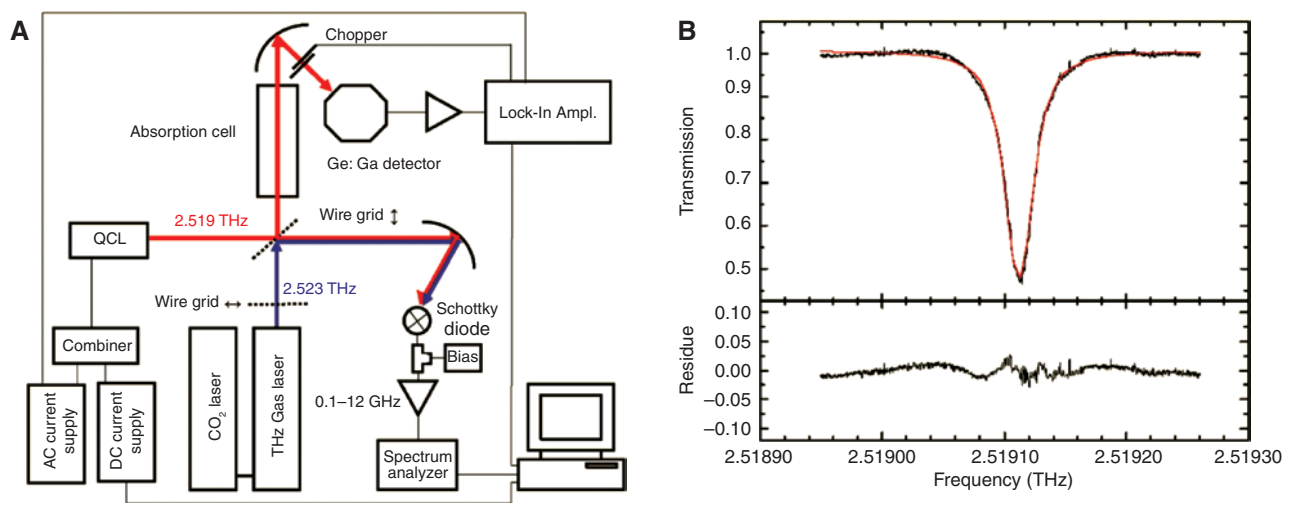

**Figure 14:** High-precision THz spectroscopy on methanol vapours with a QCL frequency referenced to a molecular transition. (A) Experimental setup and (B) acquired absorption spectrum measured at 100 Pa from [129], Copyright 2006.

In another work dating back to 2009 by Khosropanah et al., although molecular spectroscopy was not performed, the possibility of referencing a QCL to the primary frequency standard was first demonstrated by phase-locking a 2.7-THz QCL to a microwave reference [130]. However, in analogy with other well-developed spectral regions, the key tool that revolutionized the field of frequency metrology was the advent of frequency comb synthesizer (FCS). Multiple approaches have been tried to migrate the qualities of FCSs to the terahertz region; in a few cases, a link between a continuous wave (CW) terahertz source and a near-IR comb has been provided by detection techniques based on photo-conductive antennas [131] or electro-optic crystals [132]. To detect and phase-lock the beat note, both these techniques involve the CW terahertz source in a low-efficiency up-conversion process. As a consequence, the CW-source power used for the phase lock is larger than 1 mW. Moreover, although the phase-lock figures of these setups are very good, molecular spectroscopy with these phase locked devices has never been performed. Such limitations have been recently overcome by generating a free-space terahertz FCS and by directly beating it with the CW terahertz source on a square-law detector [133]. This experimental configuration allows independent optimization of the source and reference figures of merit for an efficient beat-note detection. Once an appropriate detector is used, the amount of power needed for the frequency control of the terahertz source can be dramatically reduced while making almost the whole power available for the experiment.

The principle of the THz comb beating with the THz QCL is sketched in Figure 15. Optical rectification, in a Cherenkov configuration, of a femtosecond mode-locked Ti:Sa laser occurs in a single-mode waveguide fabricated on a MgO-doped $LiNbO_3$ crystal plate. The generated radiation is a train of THz pulses, each consisting of a single electric field cycle carrying a very large spectral content (from 100 GHz up to 6 THz, centered at 1.6 THz). As the pulses are identical, the comb-like spectrum of the infinite train has a perfectly zero offset and a spacing corresponding to the 77.47 MHz repetition rate of the pump laser. A stability in the mHz range was obtained for the repetition rate, thus ensuring stability of each tooth of the THz comb at the 100-Hz level. The generation efficiency is sufficiently high to allow using a very simple setup and a commercial hot-electron-bolometer detector, with 250-MHz bandwidth, to directly observe the beating between single teeth of the comb and a small power fraction (100 nW) from the THz QCL.

To effectively stabilize the phase of the QCL emission to the frequency comb reference, a phase-lock loop (PLL) needs to be implemented. The simplified scheme of the electronic setup used for closing the PLL is given in Figure 15. The beat-note signal is mixed with a synthesized fixed frequency and processed by an analog/digital phase detector. The correction signal closes the PLL on the fast (1-MHz bandwidth) modulation input of a low-noise QCL current driver (QubeCL by ppqSense S.r.l., Campi Bisenzio, Firenze, Italy). The locked beat-note spectrum is shown in the right part of Figure 15. The electronic bandwidth of the loop is about 200 kHz, and the achieved signal-to-noise ratio (SNR) is >50 dB at 1-Hz RBW, close to the expected limit of 60 dB. By numerical integration of the beat-note spectra, we find that about 75% of the QCL power is phase-locked to the FCS emission. The phase-lock leads to a narrowing of most of the CW laser emission down to the terahertz comb tooth linewidth.

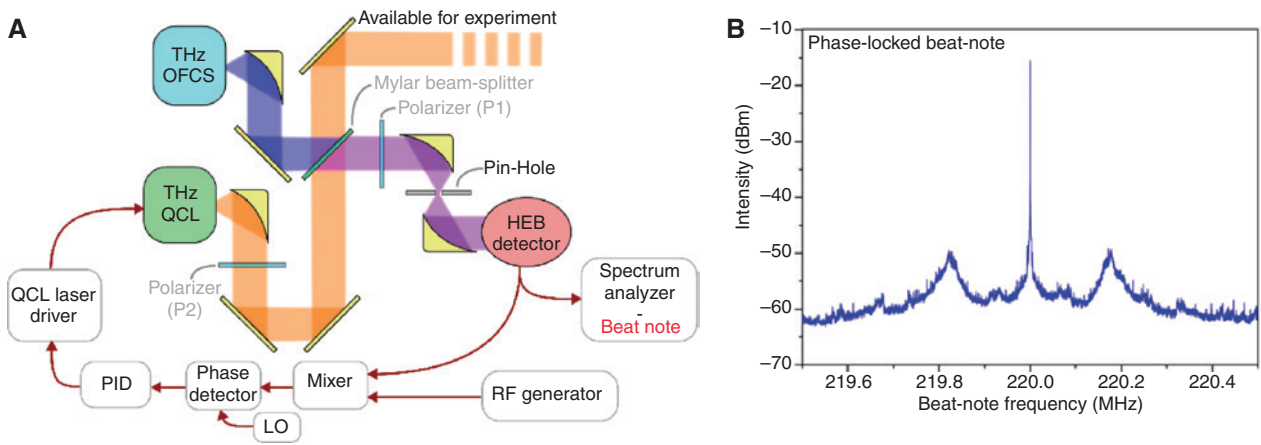

**Figure 15:** Phase locking of a THz QCL to a free-standing THz frequency comb.
Left: The beating between the QCL and the FCS is detected on a HEB. The two beams are superimposed by means of a highly asymmetrical beam splitter, so that more than 99.5% of the QCL radiation is available for the experiment beam. The P1 polarizer ensures polarization matching between the beams, whereas P2 selects the amount of QCL power to be sent to the HEB. A sketch of the electronic setup is given in the lower part of the panel. Right: beat-note spectrum when the QCL is phase-locked to the comb tooth. Reprinted with permission from [133], copyright 2012.

Exploitation of such a system for THz-comb-assisted spectroscopy has already provided precise results for rotational molecular transitions [41]. In this experiment, a direct-absorption spectroscopy setup has been implemented on a 10-cm-long cell filled with methanol gas, using the available fraction of the QCL beam (more than 99% of the total power) and an RT pyroelectric detector, together with an optical chopper on the beam and a lock-in acquisition. Once the investigated transition has been identified on a molecular database, the spectrometer can be used for accurate measurements of the absolute frequency of the identified lines. To this purpose, the line center, as well as other characteristic parameters, is determined by fitting a Voigt function to a set of experimental spectra taken at different pressures, see Figure 16. The flat residual plot shown in the bottom panel of the figure confirms the good agreement between the fitting curve and the recorded spectrum and gives a SNR higher than 200. From the SNR and the linewidth of each Voigt profile (falling in the MHz range and depending on pressure), the statistical error for the fitted line-center frequency is retrieved. It ranges from 30 to 40 kHz (depending on the data set), and it is slightly larger than the error given by the fit routine on the line center ($v_c$) parameter, thus better taking into account other error sources, such as the uncertainty on pressure. The linear dependence on pressure of the line-center frequency is shown in Figure 16B. A pressure shift of about 240 kHz/mbar is measured, and by extrapolating the $v_c$ value at zero pressure, the line-center absolute frequency is retrieved: $v_{c0} = 2553830.766(10)$ MHz. The 10-kHz error, given by the linear fit, corresponds to $4 \times 10^{-9}$ relative uncertainty, which is about two orders of magnitude worse than the accuracy of the THz comb. Indeed, the measurement is limited by the SNR of the Doppler-limited spectroscopic resolution.

Doppler-free spectroscopy using a THz QCL was first shown in ref. [134], where saturation effects could be detected using a 2.5-THz device emitting in the mW power range, even though the Lamb dip lineshape was not recorded due to radiation feedback onto the QCL. This limitation was recently overcome by Wienold and co-workers [135], proving QCL-based Doppler-free spectroscopy for the first time. The main limitation of the setup thereby presented is the lack of an absolute frequency scale, which might be soon overcome. Moreover, the demonstration of relatively high-Q resonant cavities nowadays available for THz QCLs [136] could contribute to develop other sub-Doppler spectroscopy schemes, such as cavity enhanced spectroscopy or even cavity ring down setups.

## 4.5 New generation THz QCLs

As described in the previous paragraphs, operation of CW THz QCLs is still limited at cryogenic temperatures.

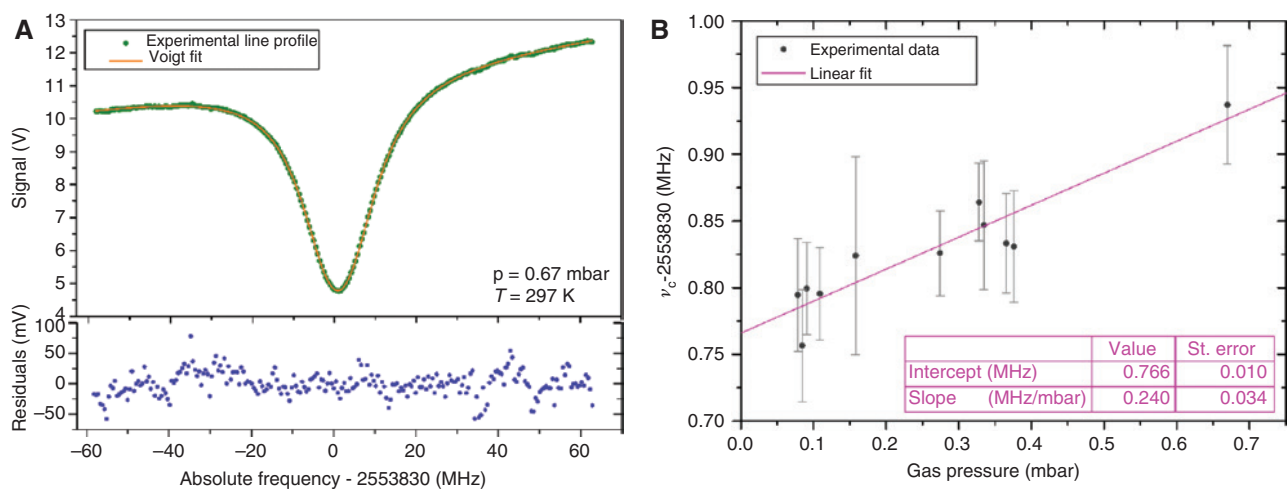

**Figure 16:** QCL-based high-precision spectroscopy on low pressure methanol vapours.
(A) Experimental absorption profile (green dots) and Voigt function fit (red line) with residuals (blue dots, bottom panel), for the investigated line. Gas pressure and temperature during the acquisition are also reported. (B) Dependence of the centerline frequency on gas pressure (black dots), with the corresponding linear fit (purple line). The intercept and the slope values of the fit give the absolute frequency ($v_{c0}$) and the pressure shift of the considered line. Reprinted with permission from [41], copyright 2014.

However, RT operation is highly desired for applications and would largely simplify any experimental setup. To address the need for RT THz sources, an alternative approach has been implemented on the basis of intracavity difference-frequency generation (DFG) in mid-IR QCLs [137–139]. With the introduction of a Cherenkov phase-matching scheme for efficient THz extraction [140, 141], THz DFG-QCLs have made marked progress in the past 5 years [142, 143]. Apart from low-resolution (~4 GHz) Fourier transform infrared (FTIR) spectrometers, in order to test whether THz DFG-QCLs are a viable alternative to THz QCLs for numerous applications that require narrow-linewidth emitters, a thorough characterization of these devices needs to be performed. These were recently presented in ref. [144].

In order to probe the linewidth of the THz emission from the device, the beat-note signal arising from the beating between the THz emission from the DFG-QCL and the free-space THz FCS was analyzed. The detected beat-note spectrum, shown in Figure 17A, can be described by a Gaussian function, whose profile was used to fit the lineshape. Because the linewidth of the THz comb tooth involved in the beating process (~130 Hz at 1 s, as experimentally demonstrated [133]) is negligible with respect to the DFG-QCL THz emission linewidth, the full width at half maximum (FWHM) of the Gaussian profiles provides an accurate quantitative estimation of the THz emission linewidth of the device. The linewidth of the 2.58-THz emission line of the CW-operated device was measured as a function of the observation time at two different temperatures (45 and 78 K), as shown in Figure 17B. The range of analysis is limited at short time scales of 20 μs. At this time scale, an upper limit of the QCL linewidth of 125 kHz was measured. Frequency referencing of this kind of sources to the primary frequency standard has not been attempted yet, but it could provide the first RT metrological grade QCL source in the near future.

Finally, in analogy with mid-IR QCLs, great efforts have been spent in the last years to replicate the operating principle of a THz FCS in a compact solid-state device, namely, a THz QCL comb, pursuing ultrashort-pulse emission in mode-locking regime. At best, these experiments have produced pulsed emission with durations of few picoseconds [145, 146], confirming the challenge of ultrashort pulsed regime. An alternative way to produce QCL-based combs relies on the development of active regions with engineered optical dispersion where, thanks to four-wave mixing, a simultaneous emission of a large number of equally spaced longitudinal modes is obtained, demonstrating also octave spanning devices [147]. However, in order to assess the true comb nature of these devices, exploitable from a metrological point of view, the equal spacing of the modes is not sufficient, while a demonstration of the presence of a well-defined, constant over time, phase relationship among all the simultaneously emitted modes has to be demonstrated.

A first insight on the modes phases of a THz QCL has been provided by shifted wave interference Fourier transform (SWIFT) spectroscopy [148, 149]. This technique can measure the phase difference between adjacent comb lines, but it cannot give a simultaneous estimation of the phase difference between modes separated by more than the detector bandwidth.

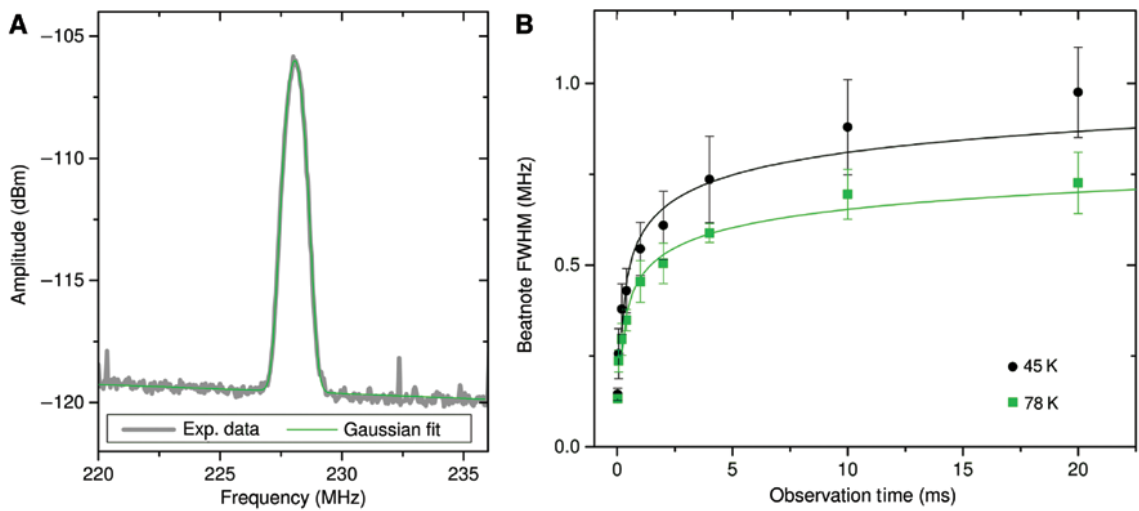

**Figure 17:** Emission linewidth of a RT THz QCLs based on intra-cavity DFG scheme.
(A) Typical beat-note spectrum observed on a spectrum analyzer for 2-ms integration time. (B) Width of the beat note at different time scales measured at two different operating temperatures of the device. The use of an FFT spectrum analyzer allows retrieval of the beat note spectra over different integration times and therefore evaluation of the THz DFG-QCL emission linewidth at different time scales (ranging from 20 μs to approximately 20 ms). Solid lines are fits using a logarithmic function. Reprinted with permission from [144], copyright 2017.

## 5 Conclusions

In conclusion, although astonishing progress has characterized research on QCLs emitting from the mid-IR to the THz range in the last 25 years, many challenges are still ahead for these key infrared sources. By summarizing the main achievements since their first demonstration in 1994 [18], the following are certainly worth mentioning: RT operation in the mid-IR [19]; extension of QCLs to the THz range [31]; experimental and theoretical demonstration of their record-low quantum-limited linewidth [43, 127, 128], among semiconductor lasers; deployment in high-resolution spectroscopic setups for precision frequency measurements [41, 47, 62]; demonstration and first characterization of direct comb emission, both in the mid-IR [83, 87, 89] and in the THz [147, 148] regions. The main challenges still ahead are many and certainly include thorough demonstration and full control of comb emission; RT operation in the THz range; full exploitation of gain media nonlinearity, which can be proven as the best way to get THz RT operation and also to generate combs by $X^2$-cascaded processes and/or for parametric generation of radiation [150]; new QCL design for significant reduction of low-frequency flicker noise, for kHz-level free-running linewidth operation [151].

**Acknowledgment:** The authors acknowledge financial support by the following:
- Ministero dell'Istruzione, dell'Università e della Ricerca Project PRIN-2015KEZNYM "NEMO – Nonlinear dynamics of optical frequency combs", Funder Id: http://dx.doi.org/10.13039/501100003407;
- European Commission H2020 Laserlab-Europe Project [654148], FET Open – ULTRAQCL Project [665158], Funder Id: http://dx.doi.org/10.13039/501100000781, "Ultrashort Pulse Generation from Terahertz Quantum Cascade Lasers" and FET Flagship on Quantum Technologies – Qombs Project [820419], "Quantum simulation and entanglement engineering in quantum cascade laser frequency combs";
- European ESFRI Roadmap "Extreme Light Infrastructure" – ELI Project.